\documentclass[10pt]{article}
\usepackage{fullpage,citesort,epsfig,psfrag,graphics,amsbsy}
\usepackage{caption}
\newcommand{\beq}{\begin{equation}}
\newcommand{\eeq}{\end{equation}}
\newcommand{\bea}{\begin{eqnarray}}
\newcommand{\eea}{\end{eqnarray}}

\newcommand{\Tr}{{\rm Tr}}
\newcommand{\be}{\begin{equation}}
\newcommand{\ee}{\end{equation}}
\newcommand{\bq}{\begin{eqnarray}}
\newcommand{\eq}{\end{eqnarray}}
\newcommand{\ket}[1]{|#1\rangle}
\newcommand{\bra}[1]{\langle#1|}

\def\math{\mathsurround=0pt }
\def\leftrightarrowfill{$\math \mathord\leftarrow \mkern-6mu 
 \cleaders\hbox{$\mkern-2mu \mathord- \mkern-2mu$}\hfill
 \mkern-6mu \mathord\rightarrow$}
\def\overleftrightarrow#1{\vbox{\ialign{##\crcr
     \leftrightarrowfill\crcr\noalign{\kern-1pt\nointerlineskip}
     $\hfil\displaystyle{#1}\hfil$\crcr}}}

\newcommand{\VEV}[1]{\langle#1\rangle}
\let\l=\lambda

 \def\bd{\begin{document}} \def\ed{\end{document}}
\def\ds{\documentstyle} \let\fr=\frac \let\bl=\bigl \let\br=\bigr
\let\Br=\Bigr \let\Bl=\Bigl
\let\bm=\bibitem
\let\na=\nabla
\let\pa=\partial \let\ov=\overline
\def\ft#1#2{{\textstyle{{\scriptstyle #1}\over {\scriptstyle #2}}}}
\def\fft#1#2{{#1 \over #2}}
\def\vp{\varphi}
\def\sst#1{{\scriptscriptstyle #1}}
\def\oneone{\rlap 1\mkern4mu{\rm l}}
\def\td{\tilde}
\def\wtd{\widetilde}
\def\dalemb#1#2{{\vbox{\hrule height .#2pt
        \hbox{\vrule width.#2pt height#1pt \kern#1pt
                \vrule width.#2pt}
        \hrule height.#2pt}}}
\def\square{\mathord{\dalemb{6.8}{7}\hbox{\hskip1pt}}}
\def\wtd{\widetilde}
\def\R{\rlap{\rm I}\mkern3mu{\rm R}}
\def\im{{\rm i}}
\def\tilg{\tilde{g}}
\def\tilF{\tilde{F}}
\def\tilA{\tilde{A}}
\def\varf{\varphi}
\def\tilf{\tilde{\phi}}
\def\tilh{\tilde{h}}
\def\rme{{\rm e}}
\def\ep{\epsilon}
\def\0{{(0)}}
\def\9{{(9)}}
\def\8{{(8)}}
\def\7{{(7)}}
\def\6{{(6)}}
\def\5{{(5)}}
\def\4{{(4)}}
\def\3{{(3)}}
\def\2{{(2)}}
\def\1{{(1)}}
\newcommand{\trace}{{\rm Tr}}
\newcommand{\ub}{\overline{U}}
\newcommand{\vb}{\overline{V}}
\newcommand{\uh}{\widehat{U}}
\newcommand{\vh}{\widehat{V}}
\newcommand{\ubh}{\overline{\widehat{U}}}
\newcommand{\vbh}{\overline{\widehat{V}}}
\newcommand{\lb}{\bar{\l}}
\newcommand{\Fb}{\overline{F}}
\newcommand{\Fh}{\widehat{F}}
\newcommand{\Fbh}{\overline{\widehat{F}}}
\newcommand{\Ab}{\overline{A}}
\newcommand{\Ah}{\widehat{A}}
\newcommand{\Abh}{\overline{\widehat{A}}}
\newcommand{\Gb}{\overline{G}}
\newcommand{\Gh}{\widehat{G}}
\newcommand{\Gbh}{\overline{\widehat{G}}}
\newcommand{\Pb}{\overline{P}}
\newcommand{\Ph}{\widehat{P}}
\newcommand{\Pbh}{\overline{\widehat{P}}}
\newcommand{\Qb}{\overline{Q}}
\newcommand{\Qh}{\widehat{Q}}
\newcommand{\Qbh}{\overline{\widehat{Q}}}
\newcommand{\Bb}{\overline{B}}
\newcommand{\Bh}{\widehat{B}}
\newcommand{\Bbh}{\overline{\widehat{B}}}
\newcommand{\fhns}{\hat{F}^{\rm (NS)}}
\newcommand{\fhrr}{\hat{F}^{\rm (RR)}}
\newcommand{\ahns}{\hat{A}^{\rm (NS)}}
\newcommand{\ahrr}{\hat{A}^{\rm (RR)}}
\newcommand{\hhrr}{\hat{H}^{\rm (RR)}}
\newcommand{\hchi}{\hat{\chi}}
\newcommand{\hphi}{\hat{\phi}}
\newcommand{\htau}{\hat{\tau}}
\newcommand{\cG}{{\cal G}}
\newcommand{\cGb}{\overline{{\cal G}}}
\newcommand{\cH}{{\cal H}}
\newcommand{\cP}{{\cal P}}
\newcommand{\cPb}{\overline{{\cal P}}}
\newcommand{\cQ}{{\cal Q}}
\newcommand{\cQb}{\overline{{\cal Q}}}
\newcommand{\cM}{{\cal M}}
\newcommand{\cN}{{\cal N}}
\newcommand{\cO}{{\cal O}}
\newcommand{\cD}{{\cal D}}
\newcommand{\cL}{{\cal L}}
\newcommand{\vpp}{\mbox{$\langle{\scriptstyle++}\rangle$}}
\newcommand{\vmp}{\mbox{$\langle{\scriptstyle-+}\rangle$}}
\newcommand{\vppp}{\mbox{$\langle{\scriptstyle+++}\rangle$}}
\newcommand{\vmpp}{\mbox{$\langle{\scriptstyle-++}\rangle$}}
\newcommand{\vpmp}{\mbox{$\langle{\scriptstyle+-+}\rangle$}}

\begin{document}
\setlength{\captionmargin}{36pt}
\begin{titlepage}
\begin{flushright}
UFIFT-HEP-08-12\\
\end{flushright}

\vskip 3cm
\begin{center}
\begin{large}
{\bf Subcritical String and Large N QCD}
\footnote{Supported in part by the Department
of Energy under Grant No. DE-FG02-97ER-41029.} 
\end{large}

\vskip 2cm
{\large 
Charles B. Thorn\footnote{E-mail  address: {\tt thorn@phys.ufl.edu}}
}
\vskip0.20cm
{\it Institute for Fundamental Theory\\
Department of Physics, University of Florida,
Gainesville FL 32611}

%(\today)

\vskip 1.0cm
\end{center}

\begin{abstract}
\noindent
We pursue the possibility of using subcritical string theory in
4 space-time dimensions to establish a string dual for
large $N$ QCD. In particular we study the even G-parity
sector of the 4 dimensional Neveu-Schwarz dual resonance model 
as the natural candidate for this string theory. Our point of
view is that the open string dynamics given by this model will
{\it determine} the appropriate subcritical closed string
theory, a tree level background of which 
should describe the sum of planar multi-loop
open string diagrams. We examine the one loop open string diagram,
which contains information about the closed string spectrum
at weak coupling. Higher loop open string
diagrams will be needed to determine closed string interactions.
We also analyze the field theory limit of the one loop open string
diagram and recover the correct running coupling behavior of the limiting
gauge theory.
\end{abstract}
\vfill
\end{titlepage}
\section{Introduction}
The underlying logic for field/string duality does not strictly
involve supersymmetry, although that symmetry plays a very important
practical role in the tractability of Maldacena's
original ${\cal N}=4$ Yang-Mills/IIB String on AdS$_5\times$S$^5$ equivalence
\cite{maldacenasole}.
This logic involves three basic facts about string theory:
\begin{enumerate}
\item The low energy limit ($\alpha^\prime\to0$)
of open string dual resonance models (DRM) is generically the tree
approximation of some flat space matrix
quantum field theory (with $SU(N)$ Chan-Paton factors, this
QFT is more specifically, a nonabelian gauge theory with gauge group
$SU(N)$).
\item The sum of planar open string DRM multi-loop diagrams has
the low energy limit of the sum of planar diagrams in the QFT, which
gives the large $N$ limit \cite{thooftlargen} of the matrix QFT.  
\item A planar open string loop can be interpreted as a tree emission
of a closed string which is absorbed into the vacuum.
\end{enumerate}
Notice that, from the closed string point of view, the sum of planar diagrams
is just a tree level shift of the vacuum. If we tried to describe the
low energy closed string dynamics 
by an effective quantum field theory, this vacuum
shift could be accomplished by solving classical field equations. In the
case of ${\cal N}=4$ such an effective field theory description is
valid in the large 't Hooft coupling limit, which is the regime that
has been most systematically studied over the last decade.

However it is not meaningful to make a strong coupling approximation
in QCD, because it is asymptotically free. Thus any attempt to apply an
effective field theory analysis to a string dual of QCD 
should be taken with a grain
of salt. It {\it might} reflect some qualitative feature of QCD,
but it could just as probably be completely misleading. Thus we expect
that even after finding the dual string theory for QCD, we will
have to deal with the vacuum shift representing the sum of planar
diagrams as a true string theory, not an effective field theory.

There has been a huge effort to adapt the
AdS/CFT paradigm to construct a string dual to QCD.
The mainstream approach to this problem has been to introduce schemes
that break the symmetries of the ${\cal N}=4$ theory down
to those of QCD. This approach was first proposed by Witten
\cite{wittenadsblackhole}, who found a way to break the supersymmetries
by replacing the AdS space on the string side with an Einstein
manifold that was a black hole embedded in AdS. 

Here we follow another path, which is to base the dual
QCD string construction on the original Neveu-Schwarz (NS) dual resonance model
in four spacetime dimensions \cite{neveuschwarz},
with all odd G-parity states projected out. For
brevity we shall call this model NS+ in this article.\footnote{
The internal consistency of this NS+ model at the
level of open strings has been appreciated at least since 
May 1971: just after Halpern and I discovered a
5 dimensional modification of the NS model with
no tachyons \cite{halpernt}, Mandelstam pointed out to us that this
NS+ model is a much simpler (indeed the simplest) 
tachyon free dual resonance model \cite{mandelstam}. 
Later I tried to stimulate interest in this model
for its own sake at the Santa Fe meeting \cite{thornsantafe}.} 
It describes an open string theory whose low energy limit has
long been known to be precisely 
Yang-Mills theory in four spacetime dimensions \cite{neveuscherk}, 
essentially because the lowest state of the open NS+ string is a massless
gauge particle. There is no open string tachyon in the
even G-parity sector. The application of this model to the
construction of a string formulation of QCD was first explicitly 
suggested by Polyakov \cite{polyakovwall}. Note that $N\to\infty$
suppresses the coupling of fields in the fundamental representation
of $SU(N)$ so that infinite N QCD is the same as infinite N pure Yang-Mills
theory. In the following when we refer to QCD, we mean this infinite N
QCD, which involves only the purely gluonic sector of QCD.

The $D=10$ version of this model has come to be known as the type-0
string model because it has no supersymmetry 
\cite{dixonharvey,seibergwitten,thornsantafe}. Since it is 
formulated in the 
critical dimension its consistent coupling to closed strings
is known. With the introduction of D3-branes, one can engineer its
low energy limit to be Yang-Mills coupled to 6 adjoint
scalar fields \cite{klebanovtseytlintype0}.

For our purposes, though, we take $D=4<10$, so we 
work directly with the subcritical string, rather than
try to embed QCD in a 10 dimensional critical string theory.
Subcritical string theory is not well understood. It is believed
that its consistent realization will involve a new scalar
(Liouville) worldsheet field which can be designed to 
cancel the conformal anomaly \cite{polyakov,curtrightthorn,gervaisneveu}.
However, we still lack a completely satisfactory 
formulation of such theories.
Recall that the unresolved issues
are associated with our imperfect understanding of the closed
string sector in subcritical theories. On the other hand 
the subcritical open string dual
models are not only self-consistent and well understood,
but they are also known to {\it imply} the existence and dynamics
of closed strings via unitarity.
We therefore adopt
the working hypothesis that the appropriate closed string
theory we seek can eventually be extracted from the open string
multi-loop diagrams \cite{thornsubdual,preitschopf}.

Incidentally, although this is not usually done, we could put the
${\cal N}=4$/AdS$_5\times$S$^5$ correspondence in this same setting.
We would first ``lift'' the ${\cal N}=4$ theory to its
simplest open string parent, in this case the Neveu-Schwarz-Ramond
open superstring \cite{neveuschwarz,ramond,thornfermi,gliozziso},
vibrating in 10 dimensions but with ends fixed to a stack of
D3-branes. Then we would ``discover'' the closed strings  and their
dynamics in the nonplanar diagrams,
and finally we would interpret the sum of multi-loop planar open string 
diagrams as a closed string background sourced by the
D3-branes. Of course, one would
still need to further recognize that the strong 't Hooft coupling
limit coupled with $\alpha^\prime\to0$ 
would validate an effective field theory determination of this
background to be AdS$_5\times$S$^5$.
%%%%%
\iffalse
There has been a long running debate through the development of
string theory of which is the ``fundamental'' starting point.
Historically open strings were the discovered first,
and very early on it was understood that closed strings and
their dynamics were derived from the consistency of open string
dynamics. From this point of view open strings are fundamental 
and closed strings are derivative. But in recent years the
consensus has been the other way around. Closed string theories 
are consistent without open strings, and moreover with
the advent of the D-brane interpretation of open strings \cite{polchinski},
the latter can be thought of as excitations of particular very special
backgrounds in closed string theory. There is also the tantalizing 
fact that closed string states typically include the graviton together
with the widespread prejudice that gravity is more fundamental than gauge
theory. But on this score there is also the idea of holography
\cite{thooftholography} which asserts that gravity is 
a manifestation of a lower dimensional fundamental theory
which has no gravity in it. A stunning example of this last
idea is the AdS/CFT correspondence. If the latter is 
actually a true duality,
one can choose either viewpoint to be the fundamental one,
depending on its practical advantages. In this paper
we start with the open string viewpoint because it
offers a clear (though not necessarily easy!) path to
the construction of the theory.
\fi
%%%%%%%%

This article initiates a program to find the subcritical closed
string theory that consistently couples to the four
dimensional even G-parity
NS open string, by analyzing the open string multi-loop
diagrams. We take the first step in this direction
by reinterpreting the 1 loop diagrams in terms of closed strings.
As observed in \cite{thornsubdual} the so-called ``unitarity
violating'' pomeron cut that arises in these diagrams can be 
interpreted as a continuous mass spectrum for the closed strings.
Alternatively one can associate this continuous mass spectrum
with a holographic fifth dimension, suggesting that the 
closed string theory we seek is best formulated in at least five 
spacetime dimensions.
The interactions between closed strings will only be revealed
in diagrams with two or more loops.

The rest of the paper is organized as follows. In Section 2
we briefly review the construction of NS tree amplitudes
for the scattering of any number of gluons. Section 3 is devoted to
a study of the 1 loop gluon amplitudes in general.
We begin by quoting the general formula for the $M$ gluon 1 loop
amplitude in the NS+ model, with enough derivation details to clearly
establish the notation and meaning of the formula. 
We also include in this section
a brief description of a very useful regularization of
these formally divergent expressions due to Goddard, Neveu,
and Scherk. Finally we discuss the closed string interpretation
including an explanation of the proper way to understand
the ``Pomeron cut''. In section 4 we analyze the field theory
limit of the one loop diagram in enough detail to extract
the renormalization group one loop beta function coefficient.
Though the coefficient for the NS+ model is the same
as the one obtained earlier for the bosonic string by Metsaev
and Tseytlin, the details of the calculation are sufficiently
different to merit a complete treatment. We close the paper
with Section 5 which contains further discussion of our results.

\section{Brief review of NS gluon tree amplitudes}
For our purposes in this article, we shall only need the 
old operator formalism of the dual resonance models. The Neveu-Schwarz
model \cite{neveuschwarz,neveust} makes use of integer moded bosonic
oscillators $a_n^\mu=(a_{-n}^\mu)^\dagger$, with
$a_0^\mu=\sqrt{2\alpha^\prime}p^\mu$, and half integer moded 
fermionic oscillators $b_r^\mu=(b_{-r}^\mu)^\dagger$
\bea
[a_n^\mu,a_m^\nu]=\eta^{\mu\nu}n\delta_{n,-m},\qquad
\{b_r^\mu,b_s^\nu\}=\eta^{\mu\nu}\delta_{r,-s}.
\eea
Here and in the following, $r,s$ will always be understood to be
half odd integers and $m,n$ to be integers. 
The string mass spectrum is given in terms of the Virasoro
generator 
\bea
L_0=\sum_{n=1}^\infty a_{-n}\cdot a_n+\sum_{r=1/2}^\infty r b_{-r}\cdot b_r
+\alpha^\prime p^2\equiv R + \alpha^\prime p^2 .
\eea 
The physical string eigenstates satisfy
\bea
(L_0-1/2)\ket{{\rm Phys}}= L_n \ket{{\rm Phys}}=G_r\ket{{\rm Phys}}=0,\qquad
n,r >0 ,
\eea
in the picture 2 Fock space \cite{neveust}. We do not need the
explicit forms for $L_n, G_r$ in this paper. G-parity in picture
2 is just $G=-(-)^{2R}$. 
Thus the even G-parity states have
the spectrum $\alpha^\prime m_e^2=-\alpha^\prime p^2=0,1,\ldots$,
and the odd G-parity states the spectrum 
$\alpha^\prime m_o^2=-\alpha^\prime p^2=-1/2,1/2,3/2,\ldots$.
The lowest mass even G-parity state is $\epsilon\cdot b_{-1/2}
\ket{0,k}$ with $k^2=0$ and $k\cdot\epsilon=0$. This massless gauge
particle state will be called the gluon in this article.

Vertex operators are constructed from the following worldsheet fields, defined
on the upper half complex plane $z=x+iy$, $y>0$:
\bea
{\cal P}(z)=\sum_n a_n z^{-n},\qquad H(z)=\sum_r b_r z^{-r},\qquad
V_0(k,z)=z^{\alpha^\prime p^2}:e^{ik\cdot (q 
+i\sqrt{2\alpha^\prime}\sum_{n\neq0}a_nz^{-n}/n)}:z^{-\alpha^\prime p^2} .
\eea
The gluon vertex operator is $g\sqrt{2/\alpha^\prime}$ times
\bea
V_\epsilon=:[\epsilon\cdot{\cal P}(1)+\sqrt{2\alpha^\prime}k\cdot H(1)
\epsilon\cdot H(1)]V_0(k,1):
\eea
and the $M$ gluon tree amplitude is then, in picture 2, the
factor
$\alpha^{\prime-1}\left(g\sqrt{2\alpha^\prime}\right)^{M-2}$
times
\bea
T_M\ =\ \bra{0,-k_1}\epsilon_1\cdot b_{1/2}V_{\epsilon_2}{1\over L_0-1/2}
V_{\epsilon_3}\cdots{1\over L_0-1/2}V_{\epsilon_{M-1}}\epsilon_M\cdot
b_{-1/2}\ket{0,k_M}. 
\eea
Note that because $k^2=k\cdot\epsilon=0$, the normal ordering of $V_\epsilon$
in the definition is not really necessary. Also notice that because
the vertex operator commutes with G-parity, the poles in $T_M$
only reveal even G-parity states: the odd G-parity states automatically
decouple in these trees. With this definition of the coupling $g$,
the QCD coupling $\alpha_s N_c\equiv g_s^2N_c/4\pi=g^2/2\pi$. 
In particular, $g$ is
held fixed in 't Hooft's $N_c\to\infty$ limit.

\section{One loop multi-gluon amplitudes in the NS+ model}
The one loop amplitudes for the Neveu-Schwarz model were first
constructed by Goddard and Waltz \cite{goddardw}, who evaluated
the planar and nonplanar 1 loop diagrams for any number of 
odd G-parity tachyons,
with vertex operator $ik\cdot H V_0(k,1)$.
The calculation is easily adapted to the gluon case by 
(1) using the gluon vertex operator and (2) projecting the
trace onto even G-parity states by inserting the projector
$P=(1+G)/2$. We do this for the planar case in some detail.
The one loop amplitude is the factor $(g\sqrt{2\alpha^\prime})^{M}$ times
\bea
{\cal M}_M&=&\int_0^1 du_1\cdots du_M\int {d^Dp\over(2\pi)^D}
\Tr V_1u_1^{R+\alpha^\prime p_0^2-3/2}\cdots
V_Mu_M^{R+\alpha^\prime p_0^2-3/2}{1-(-)^{2R}\over2}\\
&=&\int dw \int {d^Dp\over(2\pi)^D} w^{\alpha^\prime p^2-3/2}
\prod_{i=2}^M {dy_i\over y_i}\prod_i y_i^{2\alpha^\prime p\cdot k_i}
%\nonumber\\&& 
\prod_{i<j} y_j^{-2\alpha^\prime k_i\cdot k_j}
\Tr V_1(y_1)\cdots V_M(y_M)w^{R}{1-(-)^{2R}\over2} .
\eea
Recall that here all external particles are massless. The even G-parity
projection is easily handled by doing the calculation without the
projection and then subtracting from it the expression obtained
by reversing the signs of all the $w^r$ with $r$ half integral,
and dividing the difference by 2. In the following we complete the
calculation without the projector.

The integral over $p$ is easily performed:
\bea
\int {d^Dp\over(2\pi)^D}
\exp\left\{\alpha^\prime p^2\ln w+2\alpha^\prime p\cdot \sum_i k_i\ln y_i
\right\}&=&
%\nonumber\\\hskip-1.5in
\left({-1\over4\pi\alpha^\prime\ln w}\right)^{D/2}
\exp\left\{-\alpha^\prime{(\sum_i k_i\ln y_i)^2\over\ln w}
\right\} .
\eea
We also need
\bea
\int {d^Dp\over(2\pi)^D}
\exp\left\{\alpha^\prime p^2\ln w
\right\}p^{\mu_1}\cdots p^{\mu_k}&\equiv&
\left({-1\over4\pi\alpha^\prime\ln w}\right)^{D/2}
\VEV{p^{\mu_1}\cdots p^{\mu_k}},
\eea
where $\VEV{p^{\mu_1}\cdots p^{\mu_k}}$ can be evaluated with a Wick expansion
with contractions
\bea
\VEV{p^{\mu}p^{\nu}}={-\eta^{\mu\nu}\over2\alpha^\prime\ln w}.
\eea
Because $k_i^2=0$, we have 
\bea
(\sum_i k_i\ln y_i)^2&=&{1\over2}
\sum_{i\neq j}k_i\cdot k_j\left(-\ln^2{y_i\over y_j}
+\ln^2y_i+\ln^2y_j\right)
%\nonumber\\&=&
\ =\ -\sum_{i< j}k_i\cdot k_j\ln^2{y_i\over y_j} ,
\eea
so
\bea
{\cal M}_M
&=&\int {dw\over w}  
\prod_{i=2}^M {dy_i\over y_i}w^{-1/2}
\left({-1\over4\pi\alpha^\prime\ln w}\right)^{D/2}
\exp\left\{\alpha^\prime
\sum_{i< j}k_i\cdot k_j{\ln^2{y_i/y_j}\over\ln w}\right\}
\nonumber\\&& 
\hskip1in\prod_{i<j} y_j^{-2\alpha^\prime k_i\cdot k_j}
\VEV{\Tr V_1(y_1)\cdots V_M(y_M)w^{R}}.
\label{empee}
\eea
The variables $y_i$ are given by
\bea
y_1=1,\quad y_i=u_1u_2\cdots u_{i-1},\quad w=u_1u_2\cdots u_M\\
0<w<y_M<y_{M-1}<\cdots <y_2<y_1=1\\
du_1\cdots du_M={dy_2\over y_2}\cdots{dy_M\over y_M}dw .
\eea
The gluon vertex operator is $V=e^{i k\cdot x}(\epsilon\cdot{\cal P}
+\sqrt{2\alpha^\prime}k\cdot H\epsilon\cdot H)\equiv e^{i k\cdot x}
{\hat{\cal P}}$. Then
\bea
&&\hskip-.5in\VEV{\Tr V_1(y_1)\cdots V_M(y_M)w^{R}}\ =\ 
\nonumber\\&&
\VEV{{\hat{\cal P}}(y_1)\cdots{\hat{\cal P}}(y_M)}{\prod_r(1+w^r)^D
\over\prod_n(1-w^n)^D}
\prod_{i<j}\left[\left(1-{y_j\over y_i}\right)\prod_n
{\left(1-w^n{y_i\over y_j}\right)\left(1-w^n{y_j\over y_i}\right)
\over(1-w^n)^2}\right]^{2\alpha^\prime k_i\cdot k_j}\nonumber\\
&&\hskip.5in=
\VEV{{\hat{\cal P}}(y_1)\cdots{\hat{\cal P}}(y_M)}{\prod_r(1+w^r)^D
\over\prod_n(1-w^n)^D}\prod_{i<j}y_j^{2\alpha^\prime k_i\cdot k_j}
\prod_{i<j}\left[2i{\theta_1\left({1\over2i}\ln{y_i\over y_j},\sqrt{w}\right)
\over
\theta_1^\prime(0,\sqrt{w})}
\right]^{2\alpha^\prime k_i\cdot k_j} .
\eea
Here the $\VEV{\cdots}$ is a correlator of a finite number of
${\cal P}$ and $H$ worldsheet fields determined by its Wick expansion
with the following contraction rules
\bea
\VEV{{\cal P}(y_l)}&=&\sqrt{2\alpha^\prime}\sum_i k_i\left[
-{\ln(y_i/y_l)\over\ln w}+{1\over2}{y_i+y_l\over y_l-y_i}
+\sum_{n=1}^\infty\left({y_iw^n\over y_l-y_iw^n}-{y_lw^n\over y_i-y_lw^n}
\right)\right]\nonumber\\
\VEV{{\cal P}^\mu(y_i){\cal P}^\nu(y_l)}&=&\VEV{{\cal P}^\mu(y_i)}
\VEV{{\cal P}^\nu(y_l)}
+\eta^{\mu\nu}\left[
-{1\over \ln w}+{y_i y_l\over(y_i-y_l)^2}
+\sum_{n=1}^\infty\left({y_i y_lw^n\over(y_l-y_iw^n)^2}
+{y_i y_lw^n\over(y_i-y_lw^n)^2}\right)\right]\nonumber\\
\VEV{H^\mu(y_i)H^\nu(y_j)}^+&=&\eta^{\mu\nu}
\sum_r{(y_j/y_i)^r+(wy_i/y_j)^r\over
1+w^r}\nonumber\\
\VEV{H^\mu(y_i)H^\nu(y_j)}^-&=&\eta^{\mu\nu}
\sum_r{(y_j/y_i)^r-(wy_i/y_j)^r\over
1-w^r} .
\eea
The $\pm$ superscript on the $H$ contractions distinguishes the
two types of traces over the $b_r$ oscillators: for $+$ odd
and even G-parity states contribute with the same sign, whereas
for $-$ they contribute with opposite signs. In picture 2,
the difference of the two traces projects out the odd G-parity states.

The Jacobi function $\theta_1$ has the expansions
\bea
\theta_1(z,q)&=&-i\sum_{n=-\infty}^\infty q^{(n+1/2)^2}e^{(2n+1)iz}(-)^n\\
&=&2q^{1/4}\sin z\prod_{n=1}^\infty(1-q^{2n})
\prod_{n=1}^\infty(1-q^{2n}e^{2iz})
(1-q^{2n}e^{-2iz})\\
{\theta_1(z,q)\over\theta_1^\prime(0,q)}&=&\sin z\prod_{n=1}^\infty
{(1-q^{2n}e^{2iz})(1-q^{2n}e^{-2iz})\over(1-q^{2n})^2} .
\eea
Putting $q=e^{i\pi\tau}$, $\theta_1(z|\tau)\equiv\theta_1(z,q)$, the
imaginary transform reads
\bea
\theta_1(z|\tau)=i(-i\tau)^{-1/2}e^{z^2/\pi i\tau}\theta_1\left({z\over\tau}
\bigg|-{1\over\tau}\right) .
\eea
We apply this formula with $w=e^{2i\pi\tau}$
\bea
2i\exp\left({1\over2\ln w}\ln^2{y_i\over y_j}\right)
{\theta_1\left({1\over2i}\ln{y_i\over y_j}\bigg|\tau\right)\over
\theta_1^\prime(0|\tau)}={\ln w\over\pi}
{\theta_1\left({\pi\over\ln w}\ln{y_i\over y_j}\bigg|-{1\over\tau}\right)\over
\theta_1^\prime(0|-1/\tau)}\\
{\theta_1\left({\pi\over\ln w}\ln{y_i\over y_j}\bigg|-{1\over\tau}\right)\over
\theta_1^\prime(0|-1/\tau)}=\sin{\theta_{ij}\over2}\prod_{n=1}^\infty
{(1-q^{2n}e^{i\theta_{ij}})(1-q^{2n}e^{-i\theta_{ij}})\over
(1-q^{2n})^2} ,
\eea
where $\theta_i\equiv 2\pi\ln y_i/\ln w$, $\theta_{ij}=\theta_i-\theta_j$
$q=e^{-\pi i/\tau}$. Then $dy_i/y_i={\ln w\over2\pi}d\theta_i$
and $dw/w=-\ln^2wdq/2\pi^2q$. Thus
\bea
{dw\over w}{dy_2\over y_2}\cdots{dy_M\over y_M}
={-\ln w\over\pi}\left[-{\ln w\over2\pi}\right]^M
{dq\over q}d\theta_2\cdots d\theta_M .
\eea
Because all external legs are massless, we have $2\sum_{i<j}k_i\cdot k_j
=(\sum_i k_i)^2=0$ by momentum conservation. This means that
constant factors raised to this power can be dropped: the
factor $\ln w/\pi$ in the above formula can therefore be dropped 
when it is inserted into the amplitude integrand.
\bea
\VEV{{\cal P}(y_l)}
&=&{2\pi\over
-\ln w}\sqrt{2\alpha^\prime}\sum_i k_i
\left[{1\over2}\cot{\theta_{il}\over2}
+\sum_{n=1}^\infty{2q^{2n}\sin\theta_{il}\over1-2q^{2n}\cos\theta_{il}
+q^{4n}}\right]\nonumber\\
&=&{2\pi\over
-\ln w}\sqrt{2\alpha^\prime}\sum_i k_i
\left[{1\over2}\cot{\theta_{il}\over2}
+\sum_{n=1}^\infty{2q^{2n}\over1-q^{2n}}
\sin n\theta_{il}\right]\nonumber\\
&=&{2\pi\over
-\ln w}\sqrt{2\alpha^\prime}\sum_i k_i
\left[\sum_{n=1}^\infty{1+q^{2n}\over1-q^{2n}}
\sin n\theta_{il}\right]\\
\VEV{{\cal P}(y_i){\cal P}(y_l)}
-\VEV{{\cal P}(y_i)}\VEV{{\cal P}(y_l)}
&=&{4\pi^2\over \ln^2w}\left[{1\over4}\csc^2{\theta_{il}\over2}
+\sum_{n=1}^\infty{2q^{2n}(2q^{2n}-[1+q^{4n}]\cos\theta_{il})
\over(1-2q^{2n}\cos\theta_{il}
+q^{4n})^2}\right]\nonumber\\
&=&{4\pi^2\over \ln^2w}\left[{1\over4}\csc^2{\theta_{il}\over2}
-\sum_{n=1}^\infty n{2q^{2n}\over1-q^{2n}}
\cos n\theta_{il}\right]\nonumber\\
&=&{4\pi^2\over \ln^2w}\left[-\sum_{n=1}^\infty n{1+q^{2n}\over1-q^{2n}}
\cos n\theta_{il}\right]\\
\VEV{H(y_i)H(y_j)}^+
&=& -{2\pi\over\ln w}\left[{1\over2\sin(\theta_{ji}/2)}+
2\sin{\theta_{ji}\over2}\sum_{n=1}^\infty(-)^n{q^n(1+q^{2n})\over
1-2q^{2n}\cos\theta_{ji}+q^{4n}}\right]\nonumber\\
&=&{2\pi i\over\ln w}\sum_r{e^{ir\theta_{ji}}
+q^{2r}e^{-ir\theta_{ji}}\over
1+q^{2r}}\sim -{2\pi\over\ln w}\sum_r{1-q^{2r}\over
1+q^{2r}}\sin r\theta_{ji}\nonumber\\
&=&-{2\pi\over\ln w}\left[{1\over2\sin(\theta_{ji}/2)}
-2\sum_r{q^{2r}\sin r\theta_{ji}\over
1+q^{2r}}\right]\\
\VEV{H(y_i)H(y_j)}^-
&=&-{2\pi\over\ln w}\left[{\cos(\theta_{ji}/2)\over2\sin(\theta_{ji}/2)}
-2\sum_n{q^{2n}\sin n\theta_{ji}\over
1+q^{2n}}\right] .
\eea
In these expressions we have suppressed the space-time indices 
carried by the operators on the left as well as the $\eta^{\mu\nu}$
factors on the right.
Note that the first forms of each contraction show a singularity
at $\theta=0$, whereas this singular behavior is hidden in the second
forms. Since these singularities correspond to poles in the
invariants of the process, it is tempting to associate them with
one particle reducible diagrams and drop their contributions
when extracting the 1PIR contributions. This procedure actually seems
to work in the case of the bosonic string. However, for the
Neveu-Schwarz model we are considering here, some of these
apparently ``reducible'' contributions must be included
in the ``1PIR'' answer. This is because in constructing the one loop
diagrams in the picture 2 formalism, one has implicitly carried out
some integrations by parts, and rearranged what one calls
reducible and irreducible.

Notice that after the Jacobi transform the correlators all
acquire factors of $-2\pi/\ln w$ in such a way that each
contribution to $\VEV{\cdots}$ acquires the same factor
$(-2\pi/\ln w)^M$, where $M$ is the number of external
legs in the loop diagram. These factors compensate factors
from the Jacobian of the change of integration variables.
Thus
\bea
\VEV{\cdots}_{y,w}{dw\over w}{dy_2\over y_2}\cdots{dy_M\over y_M}
={-\ln w\over\pi}\VEV{\cdots}_{\theta,q}
{dq\over q}d\theta_2\cdots d\theta_M ,
\eea
where $\VEV{\cdots}_{\theta,q}$ is computed without the $-2\pi/\ln w$
factors.

The various partition functions have the following transformation properties:
\bea
w^{1/24}\prod_n(1-w^n)&=&\left(-{\ln w\over2\pi}\right)^{-1/2}q^{1/12}
\prod_n(1-q^{2n})\\
w^{-1/48}\prod_r(1+w^r)&=&q^{-1/24}\prod_r(1+q^{2r})\\
w^{1/24}\prod_n(1+w^n)&=&{1\over\sqrt{2}}q^{-1/24}
\prod_r(1-q^{2r})\\
w^{-1/48}\prod_r(1-w^r)&=&\sqrt{2}q^{1/12}\prod_n(1+q^{2n}) ,
\eea
where $n=1,2,\cdots$, $r=1/2,3/2,\cdots$. 
The partition function factor in the loop integrand is
\bea
{\prod_r(1+w^r)^{D-2}\over\prod_n(1-w^n)^{D-2}}
=w^{(D-2)/16}q^{-(D-2)/8}\left(-{\ln w\over2\pi}\right)^{(D-2)/2}
{\prod_r(1+q^{2r})^{D-2}\over\prod_n(1-q^{2n})^{D-2}}
\eea
in the critical dimension (here $D=10$) after removal of spurious states.
In projecting out the odd G-parity states we also need the partition function
with $w^r\to-w^r$:
\bea
{\prod_r(1-w^r)^{D-2}\over\prod_n(1-w^n)^{D-2}}
=w^{(D-2)/16}2^{(D-2)/2}\left(-{\ln w\over2\pi}\right)^{(D-2)/2}
{\prod_n(1+q^{2n})^{D-2}\over\prod_n(1-q^{2n})^{D-2}} ,
\eea
also in the critical dimension (here $D=10$).

%%%%%%%%%%
For $D<10$ the physical state conditions eliminate fewer states
than in the critical dimension, though all the physical
states still have positive norm \cite{goddardt}. In this case the methodology
for removing spurious states from loops is that of Brower and Thorn
\cite{browert}, adapted to the Neveu-Schwarz case in \cite{goddardw}.
In the subcritical case the null spurious states are all of the 
form $G_{-1/2}\ket{{\rm Phys},L_0=0}$. Consequently, as in
\cite{browert} the partition function power is reduced from
$D$ to $D-1$ and, because of this restricted form of the null states,
there is a further factor of $1-w^{1/2}=(1-w)/(1+w^{1/2})$.
Roughly speaking, we may say that only one component 
of $a_{n}^\mu$ and one component of
$b_{n-1/2}^\mu$ are removed when $n>1$, but two components
of both $a_{1}^\mu$ and $b_{1/2}^\mu$ are removed:
\bea
(1-w^{1/2}){\prod_r(1+w^r)^{D-1}\over\prod_n(1-w^n)^{D-1}}
&=&(1-w^{1/2})w^{(D-1)/16}q^{-(D-1)/8}%\nonumber\\&&
\left(-{\ln w\over2\pi}\right)^{(D-1)/2}
{\prod_r(1+q^{2r})^{D-1}\over\prod_n(1-q^{2n})^{D-1}} ,
\label{jacobi1}
\eea
and for $w^r\to -w^r$
\bea
(1+w^{1/2}){\prod_r(1-w^r)^{D-1}\over\prod_n(1-w^n)^{D-1}}
&=&(1+w^{1/2})w^{(D-1)/16}2^{(D-1)/2}%\nonumber\\&&
\left(-{\ln w\over2\pi}\right)^{(D-1)/2}
{\prod_n(1+q^{2n})^{D-1}\over\prod_n(1-q^{2n})^{D-1}} .
\label{jacobi2}
\eea
After the change of integration variables to $q,\theta$, the left over
factors of $w$ and $\ln w$ from Eqs~(\ref{empee}), (\ref{jacobi1}),
(\ref{jacobi2}) are as follows for $D<10$:
\bea
w^{-1/2}(1\mp w^{1/2})w^{(D-1)/16}&=&(1\mp w^{1/2})w^{(D-9)/16}\\
\left({-1\over4\pi\alpha^\prime\ln w}\right)^{D/2}
\left(-{\ln w\over2\pi}\right)^{(D-1)/2}{-\ln w\over\pi}
&=&2\left({1\over8\pi^2\alpha^\prime}\right)^{D/2}
\left(-{\ln w\over2\pi}\right)^{1/2}
\nonumber\\
&=& 2\left({1\over8\pi^2\alpha^\prime}\right)^{D/2}
\left(-{\pi\over\ln q}\right)^{1/2} .
\eea
In contrast for the critical dimension all the $w$ dependence
of these factors cancels:
\bea
w^{-1/2}w^{(D-2)/16}&=&w^{(D-10)/16}\to1\\
\left({-1\over4\pi\alpha^\prime\ln w}\right)^{D/2}
\left(-{\ln w\over2\pi}\right)^{(D-2)/2}{-\ln w\over\pi}
&=&2\left({1\over8\pi^2\alpha^\prime}\right)^{D/2}
\ \to\ 2\left({1\over8\pi^2\alpha^\prime}\right)^{5} .
\eea
Incidentally, for the subcritical bosonic string ($D<26$)
these extra factors are
\bea
w^{-1}(1-w)w^{(D-1)/24}&=&(1-w)w^{(D-25)/24}\\
\left({-1\over4\pi\alpha^\prime\ln w}\right)^{D/2}
\left(-{\ln w\over2\pi}\right)^{(D-1)/2}{-\ln w\over\pi}
&=&2\left({1\over8\pi^2\alpha^\prime}\right)^{D/2}
\left(-{\ln w\over2\pi}\right)^{1/2}\nonumber\\
&=&2\left({1\over8\pi^2\alpha^\prime}\right)^{D/2}
\left(-{\pi\over\ln q}\right)^{1/2} .
\eea
In particular, the factors of $\ln w$ work out in exactly the same way.
Of course, for $D=26$ the extra factors cancel but now $5\to13$.

Our expressions for the one loop amplitude are formal since
the integrals diverge in various regions. To give them meaning
a regularization must be found, and one should then be able
to show that divergences can be absorbed in renormalization of
parameters. Neveu and Scherk \cite{neveuscherkrenorm}, 
following an earlier suggestion of Goddard \cite{goddardreg},
showed that the divergence for $q\to1$ can be regulated by temporarily
suspending energy momentum conservation by an amount $p$:
$\sum_i k_i +p=0$. This works because essentially one is
injecting momentum $p$ into the boundary of the planar loop
with no particles attached: it can be interpreted
as the momentum of a closed string spurion. In the following
we shall refer to this procedure as the GNS regularization. It has
a very interesting feature that is illustrated by a simple
example in field theory in Appendix A. With $p\neq0$ the
two legs of an off shell propagator with a self energy insertion 
would have poles in different variables, say $p_1^2$ and $p_2^2=(p+p_1)^2$.
When the mass shift is zero, as is the case with a gauge particle,
they coalesce to a single pole as $p\to 0$, say $(Z-1)/p_1^2$. But 
then when $p\neq0$, the residues of the poles in $p_1^2$
and $p_2^2$ are each $(Z-1)/2$.
If the self energy insertion is on an external leg of an
S-matrix element, one of these legs say $p_1$ is amputated and put on shell.
If $p\neq0$ this produces a wave function renormalization factor
$(Z-1)/2$, {\it not} the $(Z-1)$ that would arise if $p=0$ from the start.
This factor of $1/2$ is precisely what is needed to end up with a properly
normalized scattering amplitude. Thus the GNS regulation is 
particularly apt for string theory amplitudes which are of necessity
always on shell. Using it, one-loop on shell diagram calculations
will automatically be correctly normalized, without the customary
$\sqrt{Z}$ adjustments that are required in usual Feynman diagram
evaluations!

In summary, we quote the one loop planar $M$ gluon NS+
amplitude for $D<10$: 
\bea
{\cal M}_M={1\over2}({\cal M}_M^+-{\cal M}_M^-)
\eea
where, in cylinder variables, $\ln q=2\pi^2/\ln w$,
\bea
{\cal M}_M^+&=&
2\left({1\over8\pi^2\alpha^\prime}\right)^{D/2}
\int \prod_{k=2}^M {d\theta_k}\int_0^1{dq\over q}\sqrt{-\pi\over\ln q}
q^{-(D-1)/8}(w^{(D-9)/16}-w^{(D-1)/16})\nonumber\\
&&\qquad
{\prod_r(1+q^{2r})^{D-1}\over\prod_n(1-q^{2n})^{D-1}}
\prod_{l<m}\left[\psi(\theta_m-\theta_l,q)
\right]^{2\alpha^\prime k_l\cdot k_m}
\VEV{{\hat{\cal P}}_1{\hat{\cal P}}_2\cdots{\hat{\cal P}}_M}^+\label{empee+}\\
{\cal M}_M^-&=&
2\left({1\over8\pi^2\alpha^\prime}\right)^{D/2}
\int \prod_{k=2}^M {d\theta_k}\int_0^1{dq\over q}\sqrt{-\pi\over\ln q}
2^{(D-1)/2}(w^{(D-9)/16}+w^{(D-1)/16})\nonumber\\
&&\qquad
{\prod_n(1+q^{2n})^{D-1}\over\prod_n(1-q^{2n})^{D-1}}
\prod_{l<m}\left[\psi(\theta_m-\theta_l,q)
\right]^{2\alpha^\prime k_l\cdot k_m}
\VEV{{\hat{\cal P}}_1{\hat{\cal P}}_2\cdots{\hat{\cal P}}_M}^-\label{empee-}\\
\psi(\theta,q)&=&\sin{\theta\over2}\prod_n{(1-q^{2n}e^{i\theta})
(1-q^{2n}e^{-i\theta})\over(1-q^{2n})^2}\nonumber\\
{\hat{\cal P}}&=& \epsilon\cdot
{\cal P}+\sqrt{2\alpha^\prime}k\cdot H\epsilon\cdot H ,
\eea
where the average $\VEV{\cdots}$ is evaluated with contractions:
\bea
\VEV{{\cal P}_l}&=&\sqrt{2\alpha^\prime}\sum_i k_i
\left[{1\over2}\cot{\theta_{il}\over2}
+\sum_{n=1}^\infty{2q^{2n}\over1-q^{2n}}
\sin n\theta_{il}\right]\\
\VEV{{\cal P}_i{\cal P}_l}
-\VEV{{\cal P}_i}\VEV{{\cal P}_l}
&=&{1\over4}\csc^2{\theta_{il}\over2}
-\sum_{n=1}^\infty n{2q^{2n}\over1-q^{2n}}
\cos n\theta_{il}\\
\VEV{H_iH_j}^+
&=&{1\over2\sin(\theta_{ji}/2)}
-2\sum_r{q^{2r}\sin r\theta_{ji}\over
1+q^{2r}}\\
\VEV{H_iH_j}^-
&=&{\cos(\theta_{ji}/2)\over2\sin(\theta_{ji}/2)}
-2\sum_n{q^{2n}\sin n\theta_{ji}\over1+q^{2n}} ,
\eea
and we have again suppressed space-time indices.
Finally  the range of integration is
\bea
0=\theta_1<\theta_2<\cdots<\theta_N<2\pi .
\eea
In these formulas $r$ ranges over positive half odd integers,
$n$ over positive integers, and $l,m \in [1,\cdots, M]$.  

\begin{figure}[ht]
\begin{center}
\includegraphics[width=2in]{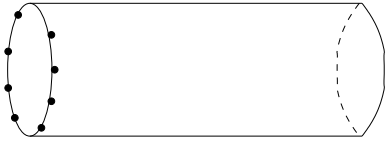}
\caption{Worldsheet of the planar loop represented as a
cylinder. The length of the cylinder is proportional to $-\ln q$.}
\label{planarws}
\end{center}
\end{figure}

\begin{figure}[ht]
\begin{center}
\includegraphics[width=2in]{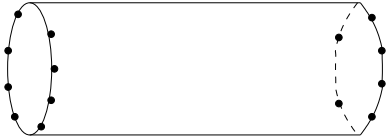}
\caption{Worldsheet for a nonplanar open string loop diagram.}
\label{nonplanarws}
\end{center}
\end{figure}

It is useful to visualize the planar loop diagram we have just quoted
as in Fig.~\ref{planarws}. 
It shows that the divergence encountered as $q\to 0$ can be interpreted
as a closed string emission into the vacuum. It also shows graphically the
physical appropriateness of the GNS regularization scheme! 
To discover the closed string spectrum one can examine the
1 loop nonplanar diagram shown in Fig.~\ref{nonplanarws}
This diagram allows the closed string to propagate with nonzero
momentum $K$. The big qualitative difference
with the planar 1-loop amplitude \cite{goddardw} is that $K^2$ now
enters the exponent of $q$ (see (\ref{empee+})):
\bea 
q^{-(D-1)/8}\to q^{-(D-1)/8+\alpha^\prime K^2/2}
\eea
so that ${\cal M}^+_{\rm NP}$ 
has a closed string cut starting at $\alpha^\prime K^2
=(D-1)/4$. Interestingly, the closed string
cut in  ${\cal M}^-_{\rm NP}$ (see (\ref{empee-})) starts instead at $K^2=0$.

As shown in \cite{thornsubdual} we can interpret the 
``unitarity violating'' closed string cut in nonplanar
diagrams as simply reflecting a continuous mass spectrum.
To see this let us rewrite the new
factors in the $D<10$ nonplanar integrand not present for 
critical dimension $D=10$:
\bea
\sqrt{-\pi\over\ln q}(w^{(D-9)/16}\mp w^{(D-1)/16})&=&
\int {d\mu\over2}q^{\mu^2/4}\left(\cosh\mu\sqrt{9-D\over16}
\mp\cos\mu\sqrt{D-1\over16}\right)\nonumber\\
&=&\int {d\mu}q^{\mu^2/4}
\cases{\sinh{\mu\gamma_+\over2}\sinh{\mu\gamma_-\over2}
&\cr
\cosh{\mu\gamma_+\over2}\cosh{\mu\gamma_-\over2}
&\cr}\\
&&\nonumber\\
\gamma_{\pm} &=& \sqrt{9-D\over16}\pm i\sqrt{D-1\over16} .
\eea
Thus we can think of the integral over $\mu$ as an integral
over a ``momentum'' in a $(D+1)$th dimension. Then the sinh (cosh)
factors can be interpreted as momentum space wave functions.
Each is a linear combination of two eigenstates of the ``position''
operator $q \equiv i\partial/\partial\mu$ with eigenvalues 
$q=\pm i\gamma_+/2$ at one
end of the cylinder and $q=\pm i\gamma_-/2$ at the other end.
Let us represent $(D+1)$th dimension by a worldsheet
scalar field $\phi$, whose zero mode is $q$. Then we see that the
one loop diagram is a sum of terms on which Dirichlet
conditions on $\phi$ are imposed: 
Open strings end on ``Dp-branes'' in $D+1$ dimensional closed
string theory, with $p=D-1$. We should therefore think of the
closed strings as propagating in the $D+1$ dimensional bulk,
and we have a holographic interpretation.
Then there is a tachyon pole at $\alpha^\prime(K^2+\mu^2)/2=(D-1)/8$ in
${\cal M}_{\rm NP}^+$, but no massless graviton poles. 
However, there are massless RR closed string poles in ${\cal M}_{\rm NP}^-$.
Specializing to $D=4$ so that the bulk is 5 dimensional
the RR tensor structures correspond to scalar, vector, and
antisymmetric tensor fields ($S,A_\mu, A_{\mu\nu}$).
We should expect that the planar diagram sum should resolve the IR issues
connected to the tachyon and the RR massless states in an interesting way.

\section{The field theory limit: Asymptotic Freedom}
For the bosonic string the field theory limit of the uv divergence
structure of the one loop diagrams has been carefully
analyzed by Metsaev and Tseytlin \cite{metsaevt}, and we follow
their logic closely. We shall specialize to the planar case,
not only for simplicity, but also because our main interest
is the relationship to large N QCD, which only includes planar
graphs. It is enough to examine the 2 and 3 gluon scattering
amplitudes to extract the one loop renormalization group coefficient.
\subsection{The two gluon function}
The two point function controls the perturbative mass shifts,
so that the two gluon function should vanish on mass shell,
because gauge particles must remain massless in perturbation theory.
Let us examine the $\theta$ integration at fixed $q$. 
First for the bosonic string, we consider the coefficient of
$\epsilon_1\cdot\epsilon_2$:
\bea
{\cal M}^{\rm Bose}_2=\int[dq]
\int_0^{2\pi} d\theta \left(\sin{\theta\over2}\prod_{n=1}^\infty
{(1-q^{2n}e^{i\theta})(1-q^{2n}e^{-i\theta})\over
(1-q^{2n})^2}\right)^{2\alpha^\prime k_1\cdot k_2}%\nonumber\\
\left[{1\over4}\csc^2{\theta\over2}
-\sum_{n=1}^\infty n{2q^{2n}\over1-q^{2n}}
\cos n\theta\right] .
\label{2gluonbose}
\eea
With no regularization, $k_2=-k_1$, $k_1^2=0$, $k_1\cdot\epsilon_1
=k_1\cdot\epsilon_2=0$, $k_1\cdot k_2=-k_1^2=0$, this expression
reduces to
\bea
\int[dq]\int_0^{2\pi} d\theta{1\over4}\csc^2{\theta\over2} , \nonumber
\eea
which is decidedly not zero. However, with the Goddard-Neveu-Scherk (GNS)
regularization, $k_2=-k_1-p$, so $2k_1\cdot k_2=(k_1+k_2)^2=p^2$, and
we have instead \cite{neveuscherkrenorm}
\bea
{1\over4}\int_0^{2\pi} d\theta \left(\sin{\theta\over2}
\right)^{\alpha^\prime p^2-2}
={1\over2}{\Gamma(1/2)\Gamma(-1/2+\alpha^\prime p^2/2)
\over\Gamma(\alpha^\prime p^2/2)}\sim -{\pi\alpha^\prime  p^2\over2}
\to 0 
\eea
as $p\to0$. Thus in the GNS regularization the gluon mass shift
is zero as it should be. Anticipating integrals done in the next section
we quote the $p\to 0$ behavior of 
\bea
{\cal M}^{\rm Bose}_2\sim \pi\alpha^\prime p^2\int[dq]
\left[-{1\over2}+4\sum_{n=1}^\infty
{q^{2n}\over (1-q^{2n})^2}\right].
\eea
We see from this
calculation that the original divergence at $\theta=0,2\pi$
was just due to the integral representation of a pole
at $\alpha^\prime p^2=1$. Since there is no pole at $p^2=0$
the analytic continuation to $p^2=0$ should be finite. The
fact that it is actually 0 is very welcome here, and is
very much due to the stringy pole structure of the
gamma functions. 

The 2 gluon function in the NS+ model is a similar
story. Since its vanishing follows from the integration over
$\theta$ the fact that the $q$ dependent factors are different
plays no role. One gets a ${\cal P}{\cal P}$ correlator whose
integral over $\theta$ gives 0 just as in the bosonic string.
The new feature is the correlator,
\bea&&\hskip-.75in 
2\alpha^\prime\VEV{k_1\cdot H\epsilon_1\cdot H
k_2\cdot H\epsilon_2\cdot H}^\pm\equiv2\alpha^\prime
(k_2\cdot\epsilon_1k_1\cdot\epsilon_2
-k_1\cdot k_2\epsilon_1\cdot\epsilon_2)C^\pm=2\alpha^\prime
\left(p\cdot\epsilon_1p\cdot\epsilon_2
-{p^2\over2}\epsilon_1\cdot\epsilon_2\right)C^\pm\nonumber\\
 C^+ &=& \left[{1\over2\sin(\theta/2)}
-2\sum_r{q^{2r}\sin r\theta\over
1+q^{2r}}\right]^2\nonumber\\
&=&
\left[{1\over4\sin^2(\theta/2)}
-{2\over\sin(\theta/2)}\sum_r{q^{2r}\sin r\theta\over
1+q^{2r}}+4\sum_{r,s}{q^{2(r+s)}\sin r\theta\sin s\theta\over
(1+q^{2r})(1+q^{2s})}\right]\\
C^- &=& 
\left[{\cos(\theta/2)\over2\sin(\theta/2)}
-2\sum_n{q^{2n}\sin n\theta\over
1+q^{2n}}\right]^2\nonumber\\
&=&
\left[{1\over4\sin^2(\theta/2)}-{1\over4}
-{2\cos(\theta/2)\over\sin(\theta/2)}\sum_n{q^{2r}\sin n\theta\over
1+q^{2n}}+4\sum_{m,n}{q^{2(m+n)}\sin m\theta\sin m\theta\over
(1+q^{2m})(1+q^{2n})}\right].
\eea 
This expression nominally vanishes as $p^2$ for $p\to0$. When
it is inserted in the integrand of the two point function, the integral over
$\theta$ of the first term in square brackets vanishes just as
in $\VEV{{\cal P}{\cal P}}$, and the integral 
of the remaining terms gives a finite contribution. Thus the $O(p^2)$
estimate for the integrand applies also for the integral over $\theta$.
Thus the new contribution in the NS+ case vanishes as $O(p^2)$
as $p\to0$. Since the coefficient of $\epsilon_1\cdot\epsilon_2$
already has an explicit $p^2$, the $p\to0$ behavior is obtained
by setting all $k_i\cdot k_j$ in the exponents to zero and using the integrals,
\bea
\int_0^{2\pi}d\theta{\sin r\theta\over \sin\theta/2}=2\pi,
\qquad \int_0^{2\pi}d\theta \sin r\theta \sin s\theta=\pi\delta_{rs},
\qquad \int_0^{2\pi}d\theta \cot(\theta/2)\sin n\theta=2\pi ,
\eea
to obtain for the new contribution to the coefficient of 
$\epsilon_1\cdot\epsilon_2$
\bea
-\pi\alpha^\prime p^2\left[-\sum_r{4q^{2r}\over1+q^{2r}}
+\sum_r{4q^{4r}\over(1+q^{2r})^2}\right]&=&
-\pi\alpha^\prime p^2\left[-4\sum_r{q^{2r}\over(1+q^{2r})^2}\right]\nonumber\\
-\pi\alpha^\prime p^2\left[-{1\over2}-\sum_n{4q^{2n}\over1+q^{2n}}
+\sum_n{4q^{4n}\over(1+q^{2n})^2}\right]&=&
-\pi\alpha^\prime p^2\left[-{1\over2}-4\sum_n{q^{2n}\over(1+q^{2n})^2}\right] .
\eea
Combining with the bose contribution gives for the Neveu-Schwarz
2 gluon function
\bea
{\cal M}^{\rm NS,+}_2&\sim& \pi\alpha^\prime p^2\int[dq]
\left[-{1\over2}+4\sum_{n=1}^\infty
{q^{2n}\over (1-q^{2n})^2}+4\sum_r{q^{2r}\over(1+q^{2r})^2}\right]\nonumber\\
{\cal M}^{\rm NS,-}_2&\sim& \pi\alpha^\prime p^2\int[dq]
\left[4\sum_{n=1}^\infty
{q^{2n}\over (1-q^{2n})^2}+4\sum_n{q^{2n}\over(1+q^{2n})^2}\right] .
\eea

\subsection{Three gluon function}
We focus here on the polarization structure $\epsilon_1\cdot\epsilon_2
\sqrt{2\alpha^\prime} k_1\cdot\epsilon_3$, which is one cyclic
ordering of the polarization structure of the 3 gluon vertex
in Yang-Mills theory.
For the bosonic string 1 loop 3 gluon
function the coefficient of this structure is
\bea
{\cal M}_3&=&\int[dq]\int_0^{2\pi}
d\theta_3\int_0^{\theta_3}d\theta_2\left[
{1\over4}\csc^2{\theta_2\over2}-\sum_{n=1}^\infty
n{2q^{2n}\over1-q^{2n}}\cos n\theta_2\right]\nonumber\\
&&\left[{\sin(\theta_2/2)\over2\sin(\theta_3/2)\sin(\theta_{32}/2)}
+\sum_{n=1}^\infty
{2q^{2n}\over1-q^{2n}}(\sin n\theta_{32}-\sin n\theta_3)\right]\nonumber\\
&&\left[\sin{\theta_2\over2}\prod_{n=1}^\infty{(1-q^{2n}e^{i\theta_2})
(1-q^{2n}e^{-i\theta_2})\over(1-q^{2n})^2}\right]^{2\alpha^\prime k_1\cdot k_2}
\left[\sin{\theta_3\over2}\prod_{n=1}^\infty{(1-q^{2n}e^{i\theta_3})
(1-q^{2n}e^{-i\theta_3})\over(1-q^{2n})^2}
\right]^{2\alpha^\prime k_1\cdot k_3}\nonumber\\
&&\left[\sin{\theta_{32}\over2}\prod_{n=1}^\infty{(1-q^{2n}e^{i\theta_{32}})
(1-q^{2n}e^{-i\theta_{32}})\over(1-q^{2n})^2}
\right]^{2\alpha^\prime k_2\cdot k_3} ,
\eea
where we include all momentum independent factors in $[dq]$.
Metsaev and Tseytlin \cite{metsaevt} extract the uv divergences in the
field theory limit by first managing the $\theta$ integrals.
They identify the 1PIR contribution by  setting the exponents
to zero and replacing the singular terms in the remaining factors by
their formal expansions
\bea
{1\over4}\csc^2{\theta\over2}&\to&-\sum_{n=1}n\cos n\theta,
\qquad {1\over2}\cot{\theta\over2}\ \to\ \sum_{n=1}^\infty\sin n\theta .
\eea
Then the $\theta$ integrals are elementary with the result
\bea
2\pi\sum_{n=1}^\infty\left({1+q^{2n}\over1-q^{2n}}\right)^2
\to 2\pi\left[-{1\over2}+4\sum_{n=1}^\infty{q^{2n}\over(1-q^{2n})^2}
\right] ,
\eea
where the formal sum $\sum_n 1$ has been interpreted as $\zeta(0)=-1/2$.

To complete the calculation we need to extract the reducible 
contributions, which we do by employing the GNS regularization of 
the string loop integral. So we introduce a spurion momentum $p$
and take $p+k_1+k_2+k_3=0$. We send $p\to0$ at the end of the calculation.
With $p\neq0$ the on-shell condition on the $k$'s allows $k_i\cdot k_j\neq0$.
Let us first extract the pole at $k_1\cdot k_2=0$, which comes from
the region $\theta_2\approx0$. Doing the integral over the region 
$0<\theta_2<\epsilon$ leads, for the $\theta$ integration, to
\bea
&&{1\over2\alpha^\prime k_1\cdot k_2}\int_0^{2\pi}d\theta_3
\left[
{1\over4}\csc^2{\theta_3\over2}-\sum_{n=1}^\infty
n{2q^{2n}\over1-q^{2n}}\cos n\theta_3\right]\nonumber\\
&&\hskip1.5in\left[\sin{\theta_3\over2}\prod_{n=1}^\infty{(1-q^{2n}e^{i\theta_3})
(1-q^{2n}e^{-i\theta_3})\over(1-q^{2n})^2}
\right]^{2\alpha^\prime (k_1+k_2)\cdot k_3}\nonumber
\eea
Defining
\bea
f(z)=\int_0^{2\pi}d\theta
\left[
{1\over4}\csc^2{\theta\over2}-\sum_{n=1}^\infty
n{2q^{2n}\over1-q^{2n}}\cos n\theta\right]
\left[\sin{\theta\over2}\prod_{n=1}^\infty{(1-q^{2n}e^{i\theta})
(1-q^{2n}e^{-i\theta})\over(1-q^{2n})^2}
\right]^z,
\eea
we are interested in its small $z$ behavior. We can expand the 
infinite product factors
\bea
\left[\prod_{n=1}^\infty{(1-q^{2n}e^{i\theta})
(1-q^{2n}e^{-i\theta})\over(1-q^{2n})^2}\right]^z
&=&1+ z \sum_{n=1}^\infty \ln{(1-q^{2n}e^{i\theta})
(1-q^{2n}e^{-i\theta})\over(1-q^{2n})^2}+O(z^2)\nonumber\\
&=&1+z\sum_{m=1}^\infty {1\over m}{2q^{2m}\over1-q^{2m}}(1-\cos m\theta)
+O(z^2) .
\eea
Working first with the contributions to the 1 term, we find
\bea
\int_0^{2\pi}d\theta{1\over4}\csc^2{\theta\over2}\left[\sin{\theta\over2}
\right]^z&=&{1\over2}{\Gamma((z-1)/2)\Gamma(1/2)\over\Gamma(z/2)}
=-{\pi\over2}z +O(z^2)\\
\int_0^{2\pi}d\theta \cos n\theta\left[\sin{\theta\over2}
\right]^z&=&-{z\over 2n}\int_0^{2\pi}d\theta {\sin n\theta\over \sin(\theta/2)}
\cos{\theta\over2}\left[\sin{\theta\over2}\right]^z\sim -{\pi\over n}z .
\eea
Thus the 1-term contribution is
\bea
1-{\rm term}\sim \pi z\left[-{1\over2}+\sum_{n=1}^\infty{2q^{2n}\over1-q^{2n}}
\right]+O(z^2) .
\eea
To find the remaining terms we use
\bea
\int_0^{2\pi} d\theta {1\over4}(1-\cos m\theta)\csc^2{\theta\over2}=\pi m\\
\int_0^{2\pi} d\theta(1-\cos m\theta)\cos n\theta=-\pi\delta_{mn} ,
\eea
to get
\bea
{\rm Remaining~terms}=\pi z\left[\sum_{n=1}^\infty{2q^{2n}\over1-q^{2n}}
+\sum_{n=1}^\infty{4q^{4n}\over(1-q^{2n})^2}\right]+O(z^2) ,
\eea
so, all together,
\bea
f(z)=\pi z\left[-{1\over2} + \sum_{n=1}^\infty{4q^{2n}\over1-q^{2n}}
+\sum_{n=1}^\infty{4q^{4n}\over(1-q^{2n})^2}\right]+O(z^2)
=\pi z\left[-{1\over2} 
+4\sum_{n=1}^\infty{q^{2n}\over(1-q^{2n})^2}\right]+O(z^2) .
\eea
Now for this contribution $z=2\alpha^\prime k_3\cdot(k_1+k_2)
=-2\alpha^\prime k_3\cdot(p+k_3)=-2\alpha^\prime p\cdot k_3$,
whereas $2k_1\cdot k_2=(k_1+k_2)^2=(p+k_3)^2=p^2+2p\cdot k_3$.
Thus the reducible contribution with pole in $k_1\cdot k_2$ is
\bea
-\pi{p\cdot k_3\over p\cdot k_3+p^2/2}\left[-{1\over2} 
+4\sum_{n=1}^\infty{q^{2n}\over(1-q^{2n})^2}\right]\to
-\pi\left[-{1\over2} 
+4\sum_{n=1}^\infty{q^{2n}\over(1-q^{2n})^2}\right] .
\eea
Notice that this is just $-1/2$ times the 1PIR contribution
found in \cite{metsaevt}, as expected for usual schemes
for wave function renormalization factors.
There are two other reducible contributions to the three
gluon amplitude associated with poles in $k_1\cdot k_3$
($\theta_3-\theta_2\approx0$) and $k_2\cdot k_3$ ($\theta_3\approx2\pi$).
But inspection of the integrand of the 1 loop three gluon
amplitude in these regions shows that these contributions
will be identical to the first. Thus the net renormalization
for the three gluon scattering amplitude
will be $(1-3/2)=-1/2$ times the 1PIR result ($(D-26)/24$) 
found in \cite{metsaevt}:
\bea
\left(1-{3\over2}\right){D-26\over24}=-{D-26\over48} .
\eea
Notice that for $D=4$ this goes to $11/24$: the factor of 11
is just the well-known 11 that occurs in the one-loop Yang-Mills 
running coupling.

We should obtain this same result in the NS+ model, but
the details of the calculation are different in a very interesting way.
The measure factors are different of course, but in the field theory limit
$w\sim0$ the difference is that the factor 
$(1-w)^{-D+2}/w\sim (D-2)+1/ w$ in the
bosonic string measure is replaced by 
$(1+\sqrt{w})^{D-2}/\sqrt{w}\sim (D-2)+1/\sqrt{w}$ in the Neveu-Schwarz
measure.
In addition the NS loop integrand involves a more complicated correlator
\bea
&&\hskip-.25in
\VEV{(\epsilon_1\cdot{\cal P}+\sqrt{2\alpha^\prime}k_1\cdot H\epsilon_1\cdot H)
(\epsilon_2\cdot{\cal P}+\sqrt{2\alpha^\prime}k_2\cdot H\epsilon_2\cdot H)
(\epsilon_3\cdot{\cal P}+\sqrt{2\alpha^\prime}k_3\cdot H\epsilon_3\cdot H)}
\nonumber\\
&=&\epsilon_1\cdot\epsilon_2\left[\VEV{{\cal P}_1{\cal P}_2}\epsilon_3
\cdot\VEV{{\cal P}_3}-2\alpha^\prime k_1\cdot k_2\VEV{H_1H_2}^2\epsilon_3
\cdot\VEV{{\cal P}_3}-(2\alpha^\prime)^{3/2} k_2\cdot k_3k_1\cdot
\epsilon_3\VEV{H_1H_2}\VEV{H_1H_3}\VEV{H_2H_3}\right.\nonumber\\
&&\left.+
(2\alpha^\prime)^{3/2} k_1\cdot k_3k_2\cdot
\epsilon_3\VEV{H_1H_2}\VEV{H_1H_3}\VEV{H_2H_3}\right]+\cdots
\eea
where $\cdots$ represents all the other polarization structures.
The first term in square brackets is identical to the correlator
encountered in the bosonic string. The remaining terms, because
of the explicit factors of $k_i\cdot k_j$ are nominally a factor
of $p$ smaller than this first term. However these factors can be cancelled
by poles arising from the $\theta$ integrals in the respective regions
$\theta_2\approx0$, $\theta_3\approx0$, or $\theta_3\approx\theta_2$.
Thus these contributions look like 1 particle reducible contributions.
We examine these contributions, after making some simplifications
valid as $p\to0$.

The poles under consideration are at most $O(p^{-1})$, so we can neglect
terms in the square bracket of $O(p^2)$ or smaller. So we can replace
$k_2\cdot\epsilon_3=-(k_1+p)\cdot\epsilon_3\to-k_1\cdot\epsilon_3$.
Furthermore we can write $k_1\cdot k_2=(k_1+k_2)^2/2=(p+k_3)^2/2
=p^2/2+k_3\cdot p\to k_3\cdot p$. Similarly $k_1\cdot k_3\to
k_2\cdot p$ and $k_2\cdot k_3\to k_1\cdot p$. Finally, we can
replace
\bea
\epsilon_3\cdot\VEV{{\cal P}_3}\to\sqrt{2\alpha^\prime}k_1\cdot\epsilon_3
\left[{1\over2}\cot{\theta_3-\theta_2\over2}
-{1\over2}\cot{\theta_3\over2}
+\sum_{n=1}^\infty{2q^{2n}(\sin n(\theta_3-\theta_2)-\sin n\theta_3)
\over1-q^{2n}}\right] .
\eea
With these simplifications we see that the $H$ terms in the square bracket
combine into a common factor $2\alpha^\prime k_3\cdot p
\epsilon_1\cdot\epsilon_2 \sqrt{2\alpha^\prime}k_1\cdot\epsilon_3$ times
\bea
&&\hskip-.4in-\left[{1\over2}\cot{\theta_3-\theta_2\over2}
-{1\over2}\cot{\theta_3\over2}
+\sum_{n=1}^\infty{2q^{2n}(\sin n(\theta_3-\theta_2)-\sin n\theta_3)
\over1-q^{2n}}\right]\left[{1\over2}\csc{\theta_2\over2}
-2\sum_r{q^{2r}\sin r\theta_2\over1+q^{2r}}\right]^2\nonumber\\
&&\hskip-.5in+\left[{1\over2}\csc{\theta_2\over2}
-2\sum_r{q^{2r}\sin r\theta_2\over1+q^{2r}}\right]
\left[{1\over2}\csc{\theta_3\over2}
-2\sum_r{q^{2r}\sin r\theta_3\over1+q^{2r}}\right]
\left[{1\over2}\csc{\theta_3-\theta_2\over2}
-2\sum_r{q^{2r}\sin r(\theta_3-\theta_2)\over1+q^{2r}}\right] .
\eea
By inspection we see that this combination of terms is not
singular as either as $\theta_3\to0$ or as $\theta_3\to\theta_2$,
so these regions of integration will not produce poles.
Moreover the $\theta_2\to0$ behavior of the first line is identical
to the corresponding limit for the bosonic string, producing
a pole whose residue is $O(p)$ and so will not compensate
the explicit $p\cdot k_3$ factor. So the only contribution that
will survive the $p\to0$ limit is the $\theta_2\approx 0$ region of the
$\theta_2$ integration of the second line:
\bea
2\alpha^\prime k_3\cdot p
\int_0^\epsilon d\theta_2{\theta_2}^{2\alpha^\prime k_3\cdot p-1}
\left[{1\over2}\csc{\theta_3\over2}
-2\sum_r{q^{2r}\sin r\theta_3\over1+q^{2r}}\right]^2\sim
\left[{1\over2}\csc{\theta_3\over2}
-2\sum_r{q^{2r}\sin r\theta_3\over1+q^{2r}}\right]^2 .
\eea
Inserting this last result into the loop integrand, we encounter the
same integral as the two gluon amplitude already evaluated, the
result being
\bea
-4\pi\sum_r{q^{2r}\over(1+q^{2r})^2}, \qquad {\rm for}~ + ,\hskip-.22in&&
\eea
which is the result for $+$ correlators of $H$ fields. Retracing the
derivation for $-$ correlators leads to the result
\bea
-{\pi\over2}-4\pi\sum_n{q^{2n}\over(1+q^{2n})^2}, \qquad {\rm for}~ -.&&
\eea
To summarize, we have identified three contributions to the $q$
integrand of the 1 loop 3 gluon scattering amplitude in the NS model. 
The $\VEV{{\cal P}^3}$ correlator produces a 1PIR contribution
\bea
I_{\cal P}^{\rm1PIR}
=2\pi\left[-{1\over2}+4\sum_n{q^{2n}\over(1-q^{2n})^2}\right] ,
\eea
and a reducible contribution which is $-3/2$ times the 1PIR piece
piece:
\bea
I_{\cal P}^{\rm1PR}
=-3\pi\left[-{1\over2}+4\sum_n{q^{2n}\over(1-q^{2n})^2}\right] .
\eea
Finally there are the contributions involving $H$ correlators
which are also reducible
\bea
I_H^{1PR+}=-4\pi\sum_r{q^{2r}\over(1+q^{2r})^2},\qquad
I_H^{1PR-}=-{\pi\over2}-4\pi\sum_n{q^{2n}\over(1+q^{2n})^2} .
\eea
Combining all the contributions together gives the simple result
\bea
I^+&=&I_{\cal P}^{\rm1PIR}+I_{\cal P}^{\rm1PR}+I_H^{1PR+}
=-\pi\left[-{1\over2}+4\sum_n{q^{2n}\over(1-q^{2n})^2}
+4\sum_r{q^{2r}\over(1+q^{2r})^2}\right]\nonumber\\
I^-&=&I_{\cal P}^{\rm1PIR}+I_{\cal P}^{\rm1PR}+I_H^{1PR-}
=-\pi\left[4\sum_n{q^{2n}\over(1-q^{2n})^2}
+4\sum_n{q^{2n}\over(1+q^{2n})^2}\right] .
\eea
As discussed in \cite{metsaevt} the field theory limit
is controlled by $w\sim0$ and there it is shown that
\bea
4\sum_n{q^{2n}\over(1-q^{2n})^2}&=&-2q{d\over dq}\sum_n\ln(1-q^{2n})\nonumber\\
&=&{1\over6}+{\ln w\over2\pi^2}+{\ln^2 w\over24\pi^2}+{\ln^2w\over\pi^2}
\sum_n\ln(1-w^n)\sim{1\over6}+{\ln w\over2\pi^2}+{\ln^2 w\over24\pi^2}+O(w) .
\eea
In a similar manner it is easily seen that
\bea
4\sum_r{q^{2r}\over(1+q^{2r})^2}&=&-2q{d\over dq}\left(\sum_n\ln{1+q^{2n}\over
1-q^{2n}}+\sum_r\ln{1-q^{2r}\over
1+q^{2r}}\right)
\nonumber\\
&=&-{\ln w\over2\pi^2}+{\ln^2w\over\pi^2}w{d\over dw}
\left(\sum_n\ln{1+w^n\over1-w^n}+\sum_r\ln{1-w^r\over1+w^r}\right)
\nonumber\\
&\sim&
-{\ln w\over2\pi^2}-w^{1/2}{\ln^2 w\over\pi^2}+O(w)\nonumber\\
4\sum_n{q^{2n}\over(1+q^{2n})^2}&=&-2q{d\over dq}
\left(\sum_r\ln{(1+q^{2r})(1-q^{2r})}-\sum_n\ln{(1-q^{2n})(1+q^{2n})}\right)
\nonumber\\
&=&-{1\over2}%%%Was -{1\over3) in v1 and in prd%%%
-{\ln w\over2\pi^2}+{\ln^2w\over\pi^2}w{d\over dw}
\left(\sum_n\ln{1+w^n\over1-w^n}+\sum_r\ln{1+w^r\over1-w^r}\right)\nonumber\\
&\sim&-{1\over2}%%%Was -{1\over3) in v1 and in prd%%%
-{\ln w\over2\pi^2}+w^{1/2}{\ln^2 w\over\pi^2}+O(w) .
\eea
[Note: In v1 of this eprint, the $-1/2$ terms in the last two
lines were erroneously written as $-1/3$. The equations are now 
correct.\footnote{I thank
Francisco Rojas for detecting this error.}]
When considering the field theory limit in the NS+ model, the details
are different from the bosonic string. Recall that in the $w\to0$
limit the factor $D-2+1/w$ in the bosonic measure changes to
$D-2+1/\sqrt{w}$ in the $+$ amplitude and to $-D+2+1/\sqrt{w}$ 
in the $-$ amplitude of the NS+ model.
The $1/w$ in the bosonic string case compensates $O(w)$
contributions in the limit of $I_{\cal P}$. However in the NS
cases we only have a $1/\sqrt{w}$ and these $O(w)$ contributions
will vanish. Thus instead of $(D-26)/24$, the $I_{\cal P}$
contribute only $(D-2)/24$ to the irreducible part and
$-3(D-2)/48=-(D-2)/16$ to the reducible part. Finally the $I^\pm_H$ 
contributes $1/2$ to the reducible part. Recall that for the NS+
amplitude we need the combination
\bea
{1\over2}\left(D-2+{1\over\sqrt{w}}\right)I^+ 
- {1\over2}\left(-D+2+{1\over\sqrt{w}}\right)I^- ,
\eea
so the ``tachyon'' singularity $1/\sqrt{w}$ cancels and the $O(w^0)$
terms from the $+$ and $-$ contributions add.
Thus the total reducible part
is $-(D-10)/16$. Of course the total contribution is
\bea
{D-2\over24}-{D-10\over16}=-{D-26\over48} .
\eea
just as for the bosonic string model. This had to be the case because 
both the bosonic string and the NS+ string go to the same
gauge theory as $\alpha^\prime\to0$.
It is mildly amusing that the reducible contribution to charge 
renormalization vanishes 
in the critical dimension for the NS model ($D=10$). To the extent that
we can associate the reducible contribution to wave function renormalization,
this would mean that there is none in the critical dimension. However 
the fact remains that there is no physically meaningful distinction
between reducible and irreducible contributions to on-shell scattering
amplitudes. Physics sees only the complete package.
\section{Discussion and Conclusion}
This article is only the beginning of a substantial program. We
have studied the one loop NS+ diagram in enough detail
to confirm that it shows the correct renormalization group properties
in the field theory limit as well as the mass spectrum of the
closed string that couples to it. Along the way we have appreciated
the great utility of the GNS regularization of string loop diagrams.

We take a few lines here to describe how the properties of the
closed string revealed so far can be consistently described by
a new (Liouville) worldsheet field $\phi$. First recall the
modification of the Virasoro generators discovered by David Fairlie 
and me independently in 1971 (see \cite{fairliethorn}). 
Here we include the easy extension to the NS super Virasoro generators:
\bea
L_n = i\alpha na^5_n+{\hat L}_n,\qquad G_r=2i\alpha rb^5_r+{\hat G}_r,
\qquad
L_0={\alpha^2\over2}+{\hat L}_0 ,
\eea
where $a^5_n, b^5_r$ are the bose and fermi oscillators associated with
a ``fifth'' (really $(D+1)$th) dimension. The hatted generators are
the usual flat space generators in $D+1$ dimensions. These modified
operators satisfy the super Virasoro algebra with 
$c=D+1+8\alpha^2$. Of course, the algebra is doubled to describe
closed strings. Vanishing of the conformal anomaly requires
$c=10$ which then determines $\alpha^2=(9-D)/8$. Applying the
on-shell condition $L_0=1/2$ then determines the ``$D$ dimensional''
mass as 
\bea{\alpha^\prime M_D^2\over4}={\alpha^2\over2}-{1\over2}
+{\alpha^\prime p_5^2\over4}+R=-{D-1\over16}+{\alpha^\prime p_5^2\over4}+R.
\eea
This shows a continuous mass spectrum starting at $M^2=-(D-1)/4$
just as revealed in the one loop calculation studied here. We
also see that the holographic 5 dimensional mass spectrum is discrete.  
It is tempting to identify $\phi$ with the free field incarnation
of the Liouville field obtained via the B\"acklund transformation. 
One further piece of information from
the one loop analysis in favor
of this interpretation is the fact that the eigenfunctions
$\sinh\gamma \mu$, $\cosh\gamma \mu$ are eigenstates of
the zero mode parity operation $\mu\to -\mu$. This restriction
was essential to the success of the quantum Ba\"cklund transformation 
constructed in \cite{braatenct}. However it would be a bit
hyperbolic to claim that these coincidences establish the 
validity of the Liouville interpretation.

In our study of the field theory limit of the one loop amplitude,
we found the Goddard-Neveu-Scherk regularization indispensable,
since it respects the proper normalization of scattering amplitudes
in on-shell perturbation theory. At a more fundamental level as
applied to the sum of planar diagrams, it simply reflects the
validity of interpreting that sum as tree emission of closed
strings into the vacuum. We think this is an interesting and
valuable insight that string theory brings to quantum field
theory.

There is clearly much work that remains to be done. We have
just scratched the surface in determining the subcritical
closed string dynamics implied by the even G-parity
4D Neveu-Schwarz model. Multiloop diagrams have
yet to be determined, let alone analyzed for their closed
string content. This is a major challenge for the immediate future.
Even at the one loop level there is more to be understood.
We have only analyzed the field theory limit
of two and three gluon amplitudes, which serve to determine 
a single renormalization group coefficient. It would
be instructive to extend the analysis, at the very least, to four gluon
amplitudes.

We already know from the one loop analysis that the closed string
spectrum includes tachyonic and massless states, which signal
a breakdown of the perturbative vacuum. Could the resolution 
of this instability explain confinement in large N QCD? We must
await the determination of the closed string effective
field theory to address this question.

\vskip14pt

\noindent\underline{Acknowledgments}: 
I should like to thank Oren Bergman, Andr\'e Neveu, and
Arkady Tseytlin for very helpful discussions. I also
thank the [Department of Energy's] Institute for Nuclear Theory 
at the University of Washington, where the research described in Section 4 
was initiated, for its hospitality and the Department of
Energy for partial support.
This research was also supported in part by the Department
of Energy under Grant No. DE-FG02-97ER-41029.

%%%%%
%%%%%%%
\appendix
\section{QFT model of GNS regularization}
Introduce a neutral scalar field $\phi$ with interaction term
\bea
-{1\over4}\Tr F_{\mu\nu}F^{\mu\nu}\lambda\phi .
\eea
Then the vertex Feynman rules are
\bea
-ig[\eta_{\mu_1\mu_2}(p_1-p_2)^{\mu_3}+\eta_{\mu_3\mu_1}(p_3-p_1)^{\mu_2}
+\eta_{\mu_2\mu_3}(p_2-p_3)^{\mu_1}],\qquad &&{\rm 3~G}\nonumber\\
i\lambda[\eta_{\mu_1\mu_2}p_1\cdot p_2-p_{1\mu_2}p_{2\mu_1}],\qquad 
&&{\rm 2~G}
-\phi\nonumber\\
-i\lambda g[\eta_{\mu_1\mu_2}(p_1-p_2)^{\mu_3}
+\eta_{\mu_3\mu_1}(p_3-p_1)^{\mu_2}
+\eta_{\mu_2\mu_3}(p_2-p_3)^{\mu_1}],\qquad &&{\rm 3~G}-\phi .
\eea
A $\phi$ insertion models a loop insertion in a string tree diagram.
So the regularized 1 loop 2 gluon function is given by the $p\to0$
limit of the 1 $\phi$ two gluon vertex ($p_1+p_2+p=0$):
\bea
i\lambda(p_1\cdot p_2\epsilon_1\epsilon_2-p_1\cdot\epsilon_2
p_2\cdot\epsilon_1){(-i)^2\over p_1^2p_2^2} .
\eea
We can take $p\to0$ by first setting $p^+,{\bf p}$ to 0, and at the same time
take the light-cone gauge $\epsilon_i^+=0$, so that $p^2=0$
and $p\cdot\epsilon_i=0$.
Then we have
\bea
{i\lambda\over2}{p_1^2+ p_2^2\over p_1^2p_2^2}\epsilon_1\epsilon_2
\to{i\lambda\over p_1^2}\epsilon_1\epsilon_2, \qquad{\rm for}~p_2\to p_1
 .
\eea
This shows that the gluon wave function renormalization is
$Z=1-\lambda$. Notice that the finite momentum $p$ has
separated the poles on the two legs of the two point function,
in such a way that if one of them is put on shell first, as
would be the case for an on-shell external leg the wave function
renormalization correction is reduced by a factor of $1/2$.
This is in fact precisely what is required by the proper
application of the reduction formalism: a factor of $\sqrt{Z}$
should be associated with each external leg! 

\end{document}